# CSSDH: An Ontology for Social Determinants of Health to Operational Continuity of Care Data Interoperability

Subhashis Das[0000−0001−9663−9009], Debashis Naskar[0000−0003−1980−0756], and Sara Rodríguez González[0000−0002−3081−5177]

BISITE Research Group, Department of Computer Sciences and Automation, University of Salamanca, Salamanca, 37007, Spain.
{subhashis, debashis, sara}@usal.es

**Abstract.** The rise of digital platforms has led to an increasing reliance on technology-driven, home-based healthcare solutions, enabling individuals to monitor their health and share information with healthcare professionals as needed. However, creating an efficient care plan management system requires more than just analyzing hospital summaries and Electronic Health Records (EHRs). Factors such as individual user needs and social determinants of health, including living conditions and the flow of healthcare information between different settings, must also be considered. Challenges in this complex healthcare network involve schema diversity (in EHRs, personal health records, etc.) and terminology diversity (e.g., ICD, SNOMED-CT) across ancillary healthcare operations. Establishing interoperability among various systems and applications is crucial, with the European Interoperability Framework (EIF) emphasizing the need for patient-centric access and control of healthcare data. In this paper, we propose an integrated ontological model, the *Common Semantic Data Model for Social Determinants of Health (CSSDH)*, by combining ISO/DIS 13940:2024 ContSys with WHO Social Determinants of Health. CSSDH aims to achieve interoperability within the Continuity of Care Network.

## 1 Introduction

The rise of digital platforms and services is helping to bridge the gap between organizational requirements and citizens' needs, particularly in health and social care. This trend underscores the value of remote monitoring and home-based solutions, driving a growing demand for digitized services. However, relying solely on Electronic Health Records (EHRs) is insufficient; understanding user needs and the social context of care is essential. Key considerations include purchasing agility, socioeconomic factors, digital literacy, and consent for accessing healthcare information. Securely accessing information across settings requires a robust healthcare management architecture, including access controls and data modeling.

According to the World Health Organization (*WHO, 2024*), poorly designed and ineffective EHRs often create a significant gap between what policymakers, health professionals, and researchers know and what they need to know to improve population health. Data from various sources, such as primary care, acute care, home care, national registries, and vital statistics systems, can be easily modified or expanded to meet the informational needs of rehabilitation.

The 2024 report, *Operational Framework for Monitoring Social Determinants of Health Equity (SDHE)*, published by the WHO [12], highlights that inequities in COVID-19 exposure, illness, and death emphasize the need to transform public health data and information systems. These systems should be equity-oriented to effectively monitor SDHE [12].

Significantly, Social Determinants of Health (SDH) are estimated to contribute to 30% to 55% of health outcomes. Exploring the relationship between SDH and outcomes is becoming increasingly important in healthcare as we strive to enhance continuous patient care. And if there is no provision to capture this information in the existing EHR system, we will not be able to monitor the patient's overall health. The recent initiative by the US Department of Health also gave emphasis to Capturing and recording SDH data under *the Healthy People 2030 initiative*.

In this paper, we proposed an Integrated Ontology-based Information Model called the Common Semantic Data Model for Social Determinants of Health (CSSDH) by combining ISO 13940: System of concepts to support continuity of care (ContSys) and WHO SDH. This model enables continuity of care data interoperability. The paper is structured as follows: related work is presented in Section 2, the overall methodology is described in Section 3, and results and evaluation, along with a discussion on future work, are covered in Section 4.

## 2   Related work and shortcoming

In the current literature, there are few initiatives aimed at formalizing a model that enables the semantic interoperability of continuity of care and incorporates social care aspects into the EHR model. For instance, the Health at Home [13] project provided a conceptually integrated framework to include social care within the patient journey. Cantor and Thorpe [1] highlighted key challenges and the need to integrate SDH data into EHR systems from a population health perspective. From an ontology engineering perspective, only two ontologies are available on BioPortal: the Social Determinants of Health Ontology (SOHO) [10] and the Social Prescribing Ontology[9], both directly related to SDH. However, they are primarily lists of terms in OWL format and do not qualify as formal ontologies that can be used as an Information Model for data integration purposes [5]. On the other hand, the SDoHo ontology[2] is overloaded with classes not exclusive to SDH, such as demographic information, age, etc. A few local initiatives aim to build universal screening for SDH alongside EHRs [11], but no final information model or publicly accessible tool exists. To our knowledge, there are no integrated formal ontological models that incorporate SDH into healthcare information models. In our previously published work[4,3], we introduced a formal ontological model based on the ContSys standard, aligned with the top-level ontology DOLCE[6]. Figure 1 shows a partial view of the ContSys ontology class visualization using the WebProtégé tool. The upper part of the figure, in green, includes DOLCE classes such as *mentalObject*, *stative*, and *event*, while the other classes are domain-specific and derived from ISO/DIS 13940:2024 ContSys. In this work, we extended our model to include concepts related to SDH.

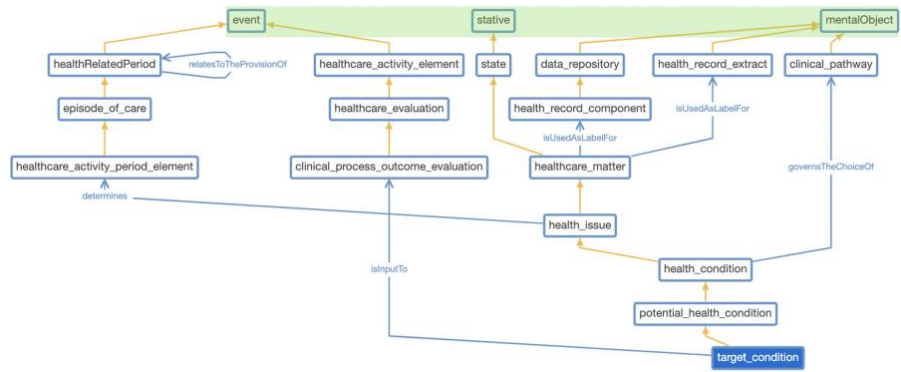

**Fig. 1.** target condition → health Condition alignment with DOLCE Top-level

## 3 Methodology

In this paper, we describe the development of a formal ontological schema for social factors influencing patients' continuous clinical outcomes, called the Common Semantic Data Model for Social Determinants of Health (CSSDH). Below, we describe the method used to develop CSSDH.

### 3.1 Modelling Concepts, Object Properties and Data Properties

In this work, we used four main resources for concept collection: the *International Classification of Diseases (ICD)-11 chapter 24*, the SOHO ontology [10], the Social Prescribing Ontology [9], and the SDH value set from the *Gravity project*. We have reused most of the terminology from the SOHO ontology and organized it based on the OntoClean methodology [8]. For example, "Lay Off From Job" within the *Economic Stability* category is modeled as a data property associated with the class *subject of care* to collect relevant information. We have extended our model with additional properties, such as "Medical Services Not Available at Home", "Problems Associated with Exposure to Radiation" and "Problems Associated with Exposure to Tobacco Smoke", borrowed from ICD-11 to make data collection more comprehensive and granular.

### 3.2    Implementation

CSSDM ontology model created using *Protégé*, an open-source ontology editing tool developed by Stanford Center for Biomedical Informatics Research. The main objective of the CSSDM model is to capture patients' SDH within the EHR system while keeping patient privacy. CSSDM should be capable of answering Competency Questions (CQ) [7] like

```
CQ1: Give me all patients lives in a low-income area.
CQ2 : Provide a list of all patients who are laid off
from their job and whose houses are also crowded.
```

CSSDM ontology has 171 classes including subject of care (i.e. patient), care professional, observation, hospital appointment, and referral. 141 relations (i.e. object properties) We have added a total of 210 data properties among them 171 are related to SDH. The rationale behind collecting SDH information only as true or false (i.e.boolean datatype) as this type of data is sensitive personal information with significant privacy and security considerations[12]. Figure 2 depicted the *subject of care* class and property restrictions on the class hierarchy panel in Protégé and on the left panel it detailed SDH-related properties (i.e. social and community factors, Neighbourhood and built-in environment) associated with the subject of care.

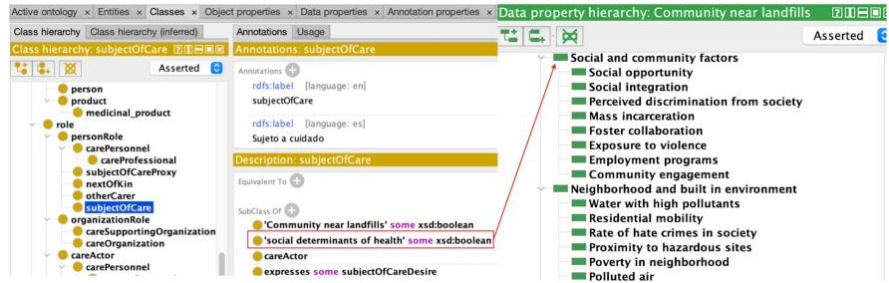

**Fig. 2.** Subject of Care SDH

## 4    Results and Discussion

Evaluation: Consistency of the CSSDM ontology is checked by the *HermiT* reasoner, a Protégé plugin. It identifies subsumption relations between classes. On the other hand Description Logic (DL) based query revealed that model is logically sound in retrieving information. For example, if policymakers would like to know *all patients who are laid off from their jobs as well as their home is crowded* (CQ2), CSSDM model can retrieve that Figure 3 depicted such an answer. We also have checked CSSDM ontology using *oops !* ontology pitfall scanner and have obtained no pitfall badge.

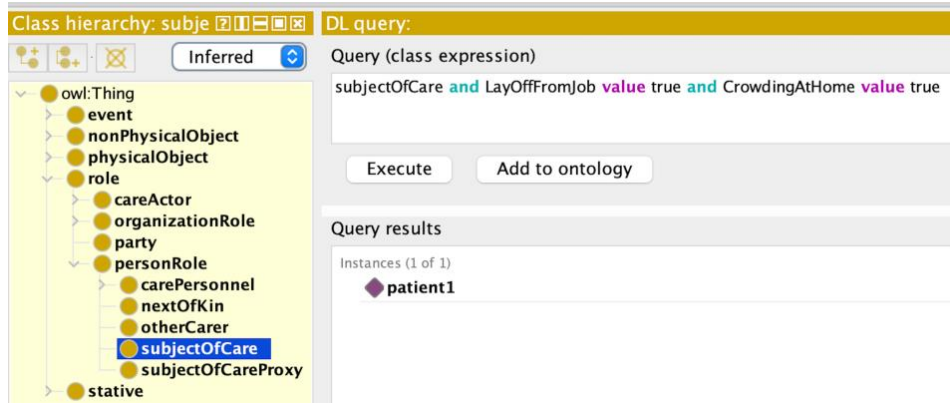

**Fig. 3.** Description Logic (DL) query execution with HermiT reasoner

We evaluated our ontology-driven CSSDM by modeling the Competency Queries elicited in Section 3.2 as SPARQL queries and assessing the returned answers. To address *CQ1*, we associated the data properties *LayOffFromJob* and *CrowdingAtHome* with the class *subjectOfCare* and set their data types as Boolean. A snapshot of the SPARQL query structure is provided.

```
PREFIX rdf: <http://www.w3.org/1999/02/22-rdf-syntax-ns#>
PREFIX owl: <http://www.w3.org/2002/07/owl#>
PREFIX rdfs: <http://www.w3.org/2000/01/rdf-schema#>
PREFIX xsd: <http://www.w3.org/2001/XMLSchema#>
PREFIX coc: <http://purl.org/net/for-coc#>
PREFIX fhir: <http://hl7.org/fhir/>
SELECT ?subjectOfcare ?forename ?Layofffromjob ?CrowdingAtHome
WHERE { ?subjectOfcare fhir:Patient.forename ?forename .
    ?subjectOfcare coc:Lay_off_from_job ?Lay_off_from_job .

OPTIONAL {?subjectOfcare coc:Crowding_at_home ?CrowdingAtHome}
}
```

*CSSDM* aims to provide a formal OWL representation of the continuity of care domain, addressing identified gaps and overcoming the limitations of loosely available SDH control vocabularies. We have extensively reused existing ontologies and standards vocabularies to facilitate patient data capture in the context of SDH. CSSDM is designed to protect patient privacy while still providing healthcare professionals with the necessary information about social factors affecting the subject of care.

We have created classes and properties that offer flexibility for data capture and integration, as well as supporting semantic reasoning. Our future plan is to test our model with a large dataset from NHS Scotland and HSE Ireland, [MIMIC-III](#) free clinical Dataset, and a regional dataset from the Recoletas Red Hospital using our existing collaboration.

Acknowledgement *This project has received funding from the European Union's Horizon 2020 research and innovation programme under the Marie Sk-lodowska-Curie grant agreement No. 101034371.*